\documentclass[aps,twocolumn]{revtex4}
\usepackage{amsfonts}
\usepackage{amsmath}
\usepackage{amssymb,epsf}
\usepackage{latexsym}
\usepackage{graphicx,epsfig}
\usepackage{amssymb}
\usepackage{float}
\usepackage{subfigure}
\usepackage{epstopdf}
\usepackage[colorlinks=true,citecolor=blue,linkcolor=blue,urlcolor=black]{hyperref}
\usepackage{dcolumn}
\usepackage{psfrag}
\usepackage{wrapfig}
\usepackage{makeidx}
\usepackage{epsf}
\usepackage{color}                                          
\usepackage{multirow}

\newcommand{\be}{\begin{equation}}
\newcommand{\ee}{\end{equation}}

\newcommand{\mincir}{\raise
-3.truept\hbox{\rlap{\hbox{$\sim$}}\raise4.truept\hbox{$<$}\ }}
\newcommand{\magcir}{\raise
-3.truept\hbox{\rlap{\hbox{$\sim$}}\raise4.truept\hbox{$>$}\ }}

\providecommand{\U}[1]{\protect\rule{.1in}{.1in}}

\begin{document}
\title{\Large Dark Universe inspired by the Kaluza-Klein gravity and impact on Primordial Gravitational Waves}
\author{Kimet Jusufi$^1$}
\email{kimet.jusufi@unite.edu.mk}
\author{Giuseppe Gaetano Luciano$^2$}
\email{giuseppegaetano.luciano@udl.cat (corresponding author)}
\author{Ahmad Sheykhi$^{3,4}$}
\email{asheykhi@shirazu.ac.ir}
\author{Daris Samart$^{5}$}
\email{darisa@kku.ac.th}
\affiliation{$^1$Physics Department,
University of Tetova, Ilinden Street nn, 1200, Tetovo, North
Macedonia}
\affiliation{$^2$Department of
Chemistry, Physics and Environmental and Soil Sciences, Escola
Politecnica Superior, Universidad de Lleida, Av. Jaume II, 69,
25001 Lleida, Spain}
\affiliation{$^3$Department of Physics,
College of
Science, Shiraz University, Shiraz 71454, Iran \\
$^4$ Biruni Observatory, College of Science, Shiraz University,
Shiraz 71454, Iran}
\affiliation{$^{5}$Khon Kaen Particle Physics and Cosmology Theory Group (KKPaCT), Department of Physics, Faculty of Science, Khon Kaen University,123 Mitraphap Rd., Khon Kaen, 40002, Thailand}

\begin{abstract}
We explore the potential implications of
Kaluza-Klein (KK) gravity in unifying the dark sector of the Universe.
Through dimensional reduction in KK gravity, the 5D spacetime framework can be reformulated in terms of a 4D spacetime metric, along with additional scalar and vector fields. 
From the 4D perspective, this
suggests the existence of a tower of particle states, including KK gravitons with spin-0 and spin-1 states, in addition to the massless spin-2 gravitons of general relativity (GR). 
The key idea in the present paper is the analogy with superconductivity theory. By assuming a minimal coupling between an additional complex scalar field and the gauge field, a "mass" term emerges for the spin-1 gravitons. This, in turn, leads to long-range gravitational effects that could modify Newton's law of gravity through Yukawa-type corrections.
Assuming an environment-dependent mass for the spin-1 graviton, near the galactic center the repulsive force from this spin-1 graviton is suppressed by an additional attractive component from Newton's constant corrections, resulting in a Newtonian-like, attraction-dominated effect. In the galaxy's outer regions, the repulsive force fades due to its short range, making dark matter appear only as an effective outcome of the dominant attractive corrections. This approach also explains dark matter's emergence as an apparent effects on cosmological scales while our model is equivalent to the scalar-vector-tensor gravity theory. 
Finally, we examine the impact of dark matter on the primordial gravitational wave (PGW) spectrum and show that it is sensitive to dark matter effects, providing an opportunity to test this theory through future GW observatories.
\end{abstract}

\maketitle

\section{Introduction}
Two main challenges of the modern cosmology are  the dark
energy puzzle \cite{Copeland:2006wr} and  the dark matter problem \cite{Arkani}. The former
originates from the observation of the accelerated expansion of Universe
in 1998 \cite{AccExp}. This discovery fundamentally challenged
modern cosmology, since the standard model of cosmology, which includes matter and radiation as two constituent components of the Universe's energy, predicts that the expansion of our Universe is decelerated due to the gravitational force between galaxies. The
accelerated Universe implies that there should be an unknown
component of energy, called dark energy, which has anti-gravity
nature and pushes our Universe to accelerate. The word \textit{dark}
here means unknown. As an alternative perspective, one can also argue that the
accelerated expansion can be understood through modification of general relativity (GR),
without the need for an additional component of energy.

On the other hand, the dark matter problem originates from the observation that the luminous (baryonic) matter in galaxies, galaxy clusters, or even the Universe at large scales does not provide sufficient gravitational force to explain the observed dynamics of these systems, necessitating the consideration of a missing component of matter to correctly account for their dynamics. Clearly, the dark matter has the
gravity nature but does not interact with electromagnetic field
and indeed this is the reason why it is called \textit{dark}.
Another approach to addressing this problem is to modify the underlying theory of gravitation (Newtonian gravity) so that the observed dynamics of the system can be naturally explained without requiring an additional component of matter.

One of the well-known alternative theories for resolving the dark matter problem through a modification of gravity is the so-called Modified Newtonian Dynamics (MOND), proposed by M. Milgrom in 1983 \cite{Milgrom1,Milgrom2,Milgrom3}. This theory appears to be highly
successful for explaining the observed anomalous
rotational-velocity. In fact, the MOND theory is an empirical
modification of Newtonian dynamics achieved by altering the
kinematic acceleration term `$a$' (which is normally expressed as
$a=v^{2}/r$) to an effective kinematic acceleration $a_{\rm eff}=a
\mu(\frac{a}{a_{0}})$, such that
\begin{equation}
\label{MOND} 
a \mu (\frac{a}{a_0})=\frac{G M}{R^2},
\end{equation}
where $ \mu=1$ for usual values of accelerations and $\mu=\frac{a}{%
a_{0}}$($\ll 1$) when the acceleration `$a$' is extremely low, below the critical value $a_{0}=10^{-10}$ m/s$^{2}$. At large
distance, at the outskirts of a galaxy, the kinematic acceleration
`$a$' is extremely small, smaller than $10^{-10}$ m/s$^{2}$,
i.e., $a\ll a_{0}$. In this regime, the function $\mu
(\frac{a}{a_{0}})=\frac{a}{a_{0}}$. Consequently, the velocity of
star in a circular orbit around the galaxy center becomes constant and does not depend on the distance. This results in a flat rotation curve, as is observed.

Let us note that although the MOND theory can explain the flat rotation curve, its theoretical origin remains unclear. Therefore, it is worth developing a gravitational theory that could naturally lead to MOND as a consequence. 
Many attempts have been
done during the past years to understand the origin of the MOND
theory from different perspectives. 
For example, in \cite{Gao,She}, it was shown that by hypothesizing that gravity is not a fundamental force but rather an entropic force arising from changes in the system's information, and by considering a modified version of this scenario known as Debye entropic force, the authors demonstrate that the origin of the MOND theory can be understood from the perspective of Debye entropic gravity. 
In \cite{Vagnozzi}, the author showed that it is possible to reproduce the MOND theory, including the explanation for flat galactic rotation curves and the Tully-Fisher relation, within the framework of mimetic gravity \cite{Mim1}, without the need for particle dark matter. In \cite{Gine:2020izd}, possible connections between MOND and modifications of the uncertainty principle at cosmological scales were explored.

Another approach to explain the flat rotation curves of spiral galaxies and reproducing the MOND theory is based on the thermodynamics-gravity conjecture. According to this conjecture, one can derive the field equations of gravity using thermodynamic arguments, such as starting from the first law of thermodynamics applied to the Universe's horizon.
The key point here is to consider an expression for horizon entropy and temperature. Any modification to the entropy will alter the field equations of gravity, such as Newton's law of gravity, the Poisson equations, the Einstein equations, and, in cosmological setups, the Friedmann equations that describe the evolution of the Universe. 

In this context, the use of the non-additive Tsallis entropy, \( S \sim A^{\beta} \), in large-scale gravitational systems has led to intriguing results. Notably, the corresponding modification to Newton’s law has been shown to account for the flat rotation curves of galaxies without the need to invoke particle dark matter. Moreover, the modified Friedmann equation derived in this framework naturally leads to an accelerated expansion of the Universe, without requiring any form of dark energy~\cite{She2}. A Lagrangian formulation of Tsallis entropy was also proposed in~\cite{DagLuc}. Given the similar power-law modification to the entropy–area relation predicted by the Barrow model~\cite{Barrow:2020tzx} - albeit motivated by entirely different considerations rooted in quantum gravity - it is, in principle, worth exploring whether similar implications might extend to Barrow cosmology as well~\cite{Saridakis:2020lrg}.
Furthermore, based on the entropic force scenario and taking the entropy associated with the horizon in the form of Kaniadakis
entropy, the authors of \cite{Gab} successfully derived the MOND
theory of gravitation (see also \cite{Drepanou:2021jiv,Luciano:2022eio,Nojiri:2019skr,Lambiase:2023ryq,Luciano:2023bai} for further approaches in these directions).

In numerous modified theories of gravity, efforts to explain the
dark sector through modified gravitational potentials have been
extensively studied. For instance, the Yukawa gravitational
potential, which emerges in various modifications of
gravity - including bimetric massive gravity \cite{Aoki:2016zgp},
Horndeski scalar-tensor theory \cite{Amendola:2019laa}, and
extended or modified $f(R)$ gravity  \cite{Capozziello:2011et} - has
gained significant attention. 
Recently, several studies \cite{J0, J1, J2, J3, J4, J5, J6} have utilized the Yukawa potential to propose an intriguing relationship on cosmological scales, linking baryonic matter, apparent dark matter, and dark energy via the expression $\Omega_{DM}(z)= \sqrt{2\, \Omega_{B,
0} \Omega_{\Lambda,  0}}{(1+z)^3}$. 
Another well-known modified gravity theory is the so-called scalar-vector-tensor gravity, proposed by Moffat (see, for example, \cite{Moffat1, Moffat2, Moffat3, Moffat4}). In this theory, the Yukawa potential can also be derived by introducing a massive vector graviton, in addition to the scalar field and the massless graviton (spin-2) particle. In this direction, we note another theoretically motivated dark electromagnetism as the origin of relativistic modified Newtonian dynamics \cite{Finster:2023mne}. 

In a recent study \cite{J4} (see also \cite{Liang:2015tfa}), the concept of modeling dark energy as a superconducting medium was proposed, where spin-2 gravitons acquire mass through their interaction with the dark energy medium. 
This results in a gravitational potential combining Yukawa and Newtonian terms, requiring both a massless and a massive spin-2 graviton. Such a theory indeed exists and is realized within the framework of bimetric theory of gravity \cite{Hassan:2011zd}. 
Unfortunately, massive gravity theories are generally inconsistent without fine-tuning. The analogy with superconductivity in \cite{J4}, however, was incomplete, lacking counterparts for the electromagnetic (spin-1) field and the scalar field, which are essential in superconductors. 
The role of the spin-1 field was entirely neglected. 

In this paper, we extend the superconductivity analogy by incorporating the spin-1 graviton as the gravitational analog of the gauge field. The natural framework to achieve this is Kaluza-Klein (KK) gravity, which inherently includes scalar and vector fields. At this stage, we would like to emphasize that, in spite of some issues with incorporating the full Standard Model gauge group, KK remains a powerful framework for exploring physics beyond GR. Indeed, by introducing extra spatial dimensions, KK theory provides a geometric foundation for gauge interactions, as seen in its simplest 5D version, where gravity naturally gives rise to an Abelian gauge field (akin to electromagnetism) and a scalar field (the dilaton). Furthermore,  its fundamental idea that gauge symmetries may emerge from higher-dimensional spacetime symmetries, has been instrumental in the development of string theory and brane-world cosmologies. Not least, KK gravity predicts the existence of massive gravitational states (KK modes), which could lead to deviations from GR at short distances and provide potential dark matter candidates. The extra-dimensional scalar fields can also influence cosmological evolution, possibly contributing to dark energy and late-time cosmic acceleration.

We explore KK theory to establish a generalized gravitational potential combining Yukawa and Newtonian terms. As we shall argue, the approach based on the superconductivity analog, where scalar field condensation gives rise to a massive spin-1 gravitons, will lead to an equivalence between KK theory and the scalar-vector-tensor theory of gravity. We also point out the role of MOND-like theory in our theory. 

The remainder of the paper is organized as follows. In the next section, we review the 5D KK theory of gravity, and then we present an analogy with the superconductivity process, where, in our case, spin-1 gravitons become massive due to the spontaneous symmetry breaking of the self-interacting scalar field.  In section \ref{EmerGal}, we study the phenomenological implications of our model and illustrate the emergence of dark matter at galactic scales. We also show how the model is equivalent to scalar-vector-tensor gravity. In section \ref{EmerCos},
we apply the theory to the cosmological setup and explore the implications of the modified Friedmann equations. In
section \ref{PGW}, we calculate the Primordial Gravitational Waves (PGW)
spectrum within the framework of the modified cosmology. We finish
with conclusions in the last section. Throughout the manuscript, we adopt natural units where $\hslash=c=k_B=1$, while retaining the gravitational constant $G$ explicitly in our expressions.

\section{Dark sector from Kaluza-Klein gravity}
\label{Dark}
Let us consider the KK gravity framework, and for simplicity, we adopt a 5D Kaluza-Klein theory with the generalized Einstein-Hilbert action  \cite{Overduin:1997sri,Pongkitivanichkul:2020txi,Waeming:2021ytf}
\begin{equation}
S_{\text{5D}} = \frac{1}{16 \pi \tilde{G}}\int d^4x dy
\sqrt{-\tilde{g}_{AB}} \tilde{R}+S_{\rm matter}\,,
\end{equation}
where $\tilde{G}$ is the five-dimensional gravitational constant and $\tilde{R}=\tilde{g}_{AB} \tilde{R}^{AB}$ the five-dimensional Ricci scalar, and $S_{\rm matter}$ gives the contribution of matter fields. The metric can be written in terms of a 4D spacetime metric, a real scalar field $\Phi$ and a gauge field $A_{\mu}$, namely (see, \cite{Overduin:1997sri,Pongkitivanichkul:2020txi,Waeming:2021ytf})
\begin{equation}
  \tilde{g}_{AB} =\left( \begin{matrix}
g_{\mu\nu} + \Phi^2 A_\mu A_\nu & \Phi^2 A_\mu \\[2mm]
\Phi^2 A_\nu & \Phi^2
\end{matrix}\right).
\end{equation}

It is known that after performing dimensional reduction in the case $D=5$, we get a theory similar to the scalar-tensor theory with an additional gauge field, i.e. \cite{Overduin:1997sri,Pongkitivanichkul:2020txi,Waeming:2021ytf}
\begin{equation}\notag
    S_{\text{4D}} = \int d^4x\,\sqrt{-g} \left( \frac{R \Phi}{16 \pi G_N}
- \frac{1}{4} \Phi^3 F_{\mu\nu} F^{\mu\nu}
- \frac{2}{3} \frac{\partial_\mu \Phi\partial^\mu \Phi}{\Phi} \right)
\end{equation}
\begin{equation}
    +\int d^4x\,\sqrt{-g} \left( 
-\frac{1}{2}\partial_\mu \tilde{\phi}\,\partial^\mu \tilde{\phi}-V(\tilde{\phi})\right),
\end{equation}
where for the matter field $S_{\rm matter}$ we have added a contribution coming from a complex scalar field $\tilde{\phi}$ with a self-interaction potential $V(\tilde{\phi})$, for reasons that will be explained below. Now by renaming $G=G_N/\Phi$,  $F_{\mu \nu}=\Phi^{-3/2} \tilde{F}_{\mu\nu}$, and $\mathcal{G}= 4 \sqrt{\Phi}/\sqrt{3}$, we are led to
\begin{equation}\notag
    S_{\text{4D}} = \int d^4x\,\sqrt{-g} \left( \frac{R}{16 \pi G}
- \frac{1}{4} \tilde{F}^{\mu \nu }\tilde{F}_{\mu
\nu }
- \frac{1}{2} \partial_\mu \mathcal{G}\,\partial^\mu \mathcal{G}\right)
\end{equation}
\begin{equation}
    +\int d^4x\,\sqrt{-g} \left( -
\frac{1}{2}\partial_\mu\tilde{\phi}\,\partial^\mu \tilde{\phi}-V(\tilde{\phi})\right),
\end{equation}
where we see that the Newton's constant in this model is not a constant but it is a running scalar field. Further one can assume that $\mathcal{G}$ and $\Phi$ are slowly varying fields and we can neglect the third term, hence we are left with
\begin{eqnarray}\notag
    S_{\text{4D}} &=& \int d^4x\,\sqrt{-g} \left( \frac{R}{16 \pi G}
- \frac{1}{4} \tilde{F}^{\mu \nu }\tilde{F}_{\mu
\nu }
\right)\\
&+&\int d^4x\,\sqrt{-g} \left( 
- \frac{1}{2}\partial_\mu \tilde{\phi}\partial^\mu \tilde{\phi}-V(\tilde{\phi})\right).
\end{eqnarray}
In the weak limit approximation we can assume the following form of perturbation in scalar field 
\begin{eqnarray}
    \Phi=\Phi_0+\delta \Phi,
\end{eqnarray}
leading to
\begin{eqnarray}
    G \simeq \frac{G_N}{\Phi_0}\left(1- \frac{\delta \Phi}{\Phi_0}\right)\equiv G_N (1+\alpha),
\end{eqnarray}
but since we expect $G=G_N$ when $\delta \Phi=0$, we can fix $\Phi_0=1$ and, further $\alpha=-\delta \Phi$, with $\alpha$ being a parameter. 
\\

The key idea in the present paper is to put forward the analogy with the superconductivity process, where  photons becomes massive under the spontaneous symmetry breaking of the self-interacting complex scalar field. In order to show this analogy, we need to couple the gauge field to the complex scalar field $\tilde{\phi}$ through the minimal coupling procedure which replaces the partial derivative  with the covariant derivative via
\begin{eqnarray}
    D_{\mu }=\partial _{\mu }-ig {A}_{\mu },
\end{eqnarray}
where $g$ is the gauge coupling constant and $\tilde{\phi} $ is a complex scalar. In our case we get
\begin{equation}
\mathcal{L} = \frac{R}{16 \pi G} - \frac{1}{2}( D^{\mu }\tilde{\phi}) ^{\ast }( D_{\mu }\tilde{\phi})-\dfrac{1}{4} \tilde{F}^{\mu \nu }\tilde{F}_{\mu
\nu }-V(\tilde{\phi}).
\label{unbroken-L}
\end{equation}%
In order to obtain the mass of the vector gauge field with the superconductivity analogy, we will employ the Anderson-Higgs mechanism by rewriting the complex scalar field $\tilde{\phi}$ as \cite{Rubakov:2002fi,poniatowski2019}
\begin{equation}
\tilde{\phi} = \tilde{\phi}_0e^{i\chi},
\label{orderparameter}
\end{equation}
where $\tilde\phi_0$ is the non-vanishing vacuum expectation value. The covariant derivative on the complex scalar $\tilde\phi$ is given by $D_\mu\tilde\phi = i\big(\partial_\mu\chi - g {A}_\mu \big)\tilde\phi$. 

Then, the Lagrangian in Eq. (\ref{unbroken-L}) becomes
\begin{equation}
\mathcal{L} = \frac{R}{16 \pi G} -\dfrac{1}{4} \tilde{F}^{\mu \nu }\tilde{F}_{\mu
\nu } - \frac{1}{2}\big| \phi_0\big|^2\left(\partial_\mu\chi -g{A}_\mu \right)^2-V(\tilde{\phi})\,.
\label{unbroken-L2}
\end{equation}
In addition, we introduce the variable gauge boson $\tilde{A}_\mu$ to ensure that the above Lagrangian is gauge invariant,
\begin{equation}
\tilde{A}_\mu = A_\mu - \frac{1}{g}\partial_\mu\chi .
\label{new-gauge-var}
\end{equation}

By applying a new gauge transformation in Eq. (\ref{new-gauge-var}) into the Lagrangian in Eq. (\ref{unbroken-L2}), one finds 
\begin{equation}
\mathcal{L} =\frac{R}{16 \pi G} -\dfrac{1}{4} \tilde{F}^{\mu \nu }\tilde{F}_{\mu
\nu }  - \frac{1}{2}\mu^2 \tilde{A}^{\mu }\tilde{A}_{\mu }-V(\tilde{\phi}),
\end{equation}%
where $\mu^2 \equiv g^2|\tilde\phi_0|^2$. 

We can now count the number of degrees of freedom. A massless graviton in 5D has five degrees of freedom given by
\begin{eqnarray}
    D\,(D-3)/2|_{D=5}=5,
\end{eqnarray}
plus of course we need to add the two degrees of freedom from the complex scalar field $\tilde\phi$. Then, via the dimensional reduction, we get two degrees of freedom corresponding to the massless spin-2 graviton in 4D 
\begin{eqnarray}
    D\,(D-3)/2|_{D=4}=2,
\end{eqnarray}
along with two degrees of freedom for the massless spin-1 graviton and one degree of freedom for the scalar graviton encoded in the real scalar field $\Phi$. As we shall see, this argument leads to the existence of a massive spin-1 graviton with three degrees of freedom, as part of the lower-dimensional effective theory. Namely, by imposing $U(1)$ symmetry breaking in superconductors, one degree of freedom of scalar field $\tilde\phi$ becomes the longitudinal part of the spin-1. We then have, for the gravity sector, two degrees of freedom for the massless spin-2 graviton, three degrees of freedom for the massive spin-1 graviton, and one degree of freedom for the spin-0 graviton. To make the analogy with superconductivity more precise, we must include the final degree of freedom from the scalar field $\tilde{\phi}$, which can play the role of a dark energy particle. The superconductivity analogy, therefore, predicts a spin-0 scalar particle that could be a hint of new physics.

Using the Anderson-Higgs mechanism of the spontaneous symmetry breaking of the superconductivity framework, we finally obtain the mass of the gauge boson term $\mu^2 \tilde{A}_\mu\tilde{A}^\mu/2$ in the KK gravity Lagrangian. It is worth noting that one degree of freedom from the complex scalar field $\tilde{\phi}=\tilde{\phi}_0 e^{i\chi}$ is ``eaten"  by the gauge boson, leading to the massive one (the longitudinal component of the massive gauge boson). As we shall argue in the next section, these spin-1 gravitons particles could gravitationally couple with baryonic matter and modify the law of gravity. 
Note that the mass term of the gauge boson $\mu$ is determined by the vacuum expectation value of the scalar field, while  $V(\tilde{\phi})$ encodes the self-interacting effect of the scalar field. 

Variation of the gravitational Lagrangian with respect to the metric gives the Einstein field equations
\begin{equation}
G_{\mu \nu}= 8 \pi G \left(T_{\mu \nu}^{\rm S}+ T_{\mu \nu}^{\rm V}+T_{\mu \nu}^{\rm M}\right),
\end{equation}
where
\begin{equation}
T_{\mu \nu}^{\rm S}=
-g_{\mu \nu} V(\tilde{\phi}),
\end{equation}
and
\begin{equation}
T_{\mu \nu}^{\rm V} = \tilde{F}_{\mu \sigma} {\tilde{F}_\nu}^\sigma-\frac{1}{4} g_{\mu \nu} \tilde{F}^2+\mu^2 \left(\tilde{A}_\mu \tilde{A}_\nu - \frac{1}{2}g_{\mu \nu} \tilde{A}_\sigma \tilde{A}^\sigma \right)
\end{equation}
Note that the vanishing vacuum expectation value $\tilde{\phi}_0$ minimizes the potential, however, there is no reason to believe that the potential must necessarily be zero.
By using the ansatz in Eq. \eqref{orderparameter}, in fact, we get the constant term $V\big(|\tilde{\phi}|^2\big)= V\big(\tilde{\phi}_0^2\big)=V_0$ that looks like the cosmological constant $\Lambda$. Thus, we get the Einstein field equations
\begin{eqnarray}
 G_{\mu \nu}+\Lambda g_{\mu \nu} =8 \pi G \left(T_{\mu \nu}^{\rm M}+T_{\mu \nu}^{\rm V}\right)\,.
\end{eqnarray}
Another point to be mention is that in the strong-gravity regime, we expect the coupling between the vector field and the spacetime background geometry to play important role, leading to a correction term in the energy-momentum tensor i.e., $T^{\rm correc.}_{\mu \nu}(g_{\mu \nu}, A_{\mu})$. However, since this work focuses on the weak-gravity regime, we neglect such contributions.

Finally, we have the equation that governs the motion of vector field. From the Lagrangian, we get the Proca equation with a mass term
\begin{equation}
\nabla_\mu \tilde{F}^{\mu\nu}-\mu^2 \tilde{A}^\mu = 0.
\end{equation}%
By imposing the gauge transform in Eq. (\ref{new-gauge-var}) 
with the Lorenz gauge $ \nabla_{\mu}\tilde{A}^{\mu}=0$, one can get the wave equation in terms for $\tilde{A}^{\mu}$ as follows
\begin{eqnarray}\label{waveequation}
    \left(\Box-\mu^2 \right) \tilde{A}^{\mu}=0.
\end{eqnarray}
We shall refer to this particle as a massive spin-1 dark graviton and proceed to explore some of its implications in the following sections.

\section{Emergence of dark matter in galactic scales}
\label{EmerGal}
In this section, we first consider the coupling between the massive gauge boson spin-1 and the galaxy’s baryonic matter fields via the following interacting Lagrangian
\begin{equation}
\mathcal{L}_I = \sqrt{\alpha_B G_N} J^\mu\tilde{A}_\mu,\quad J^\mu = M\bar\psi\gamma^\mu\psi,
\label{gauge-baryon-coupling}
\end{equation}
where $\alpha_B>0$ is a dimensionless coupling parameter due to the massive gauge boson effect. We would like to emphasize that this is a free parameter of the model, which should ultimately be
constrained through comparison with observational data (see Sec. \ref{PGW} for more details). It is also plausible
that this parameter may vary depending on the specific observational probe or
matter distribution considered.
Additionally,  $\psi$ is a baryonic field and $M$ gives the total baryonic mass enclosed in some region. 
The gravitational force between the baryonic matter fields from the gauge boson exchange in our model can be described in terms of the transition (scattering) amplitudes $\mathcal{T}_{\text{YU}}$, which is obtained from the interaction Lagrangian in Eq. (\ref{gauge-baryon-coupling}) as
\begin{equation}
\mathcal{T}_{\text{YU}} (\vec q\,) \approx \alpha_B G_N M^2\frac{1}{\vec{q}^{\,2} - \mu^2 },  
\end{equation}
where $\vec q$ is the momentum exchange between baryons. In the above amplitudes, we have used the
the approximation of heavy mass and very low momentum (velocity) of the baryonic matter with fixed spin orientation, i.e., $E=\sqrt{\vec{p}^{\,2} + M^2}\approx M$. 

Basically, the transition amplitude can be used to identify the gravitational potential energy, i.e., $\mathcal{T}_{\text{YU}}(\vec q\,) \equiv U_{\text{YU}}(\vec q\,)$, as done for the nucleon-nucleon Yukawa potential from the pion exchange interaction picture in standard quantum field theory. By applying a proper mass dimension rescaling, consistent with the Poisson equation of gravity, we obtain
\begin{equation}
\tilde\Phi_{\text{YU}}(\vec q\,) \equiv U_{\text{YU}}(\vec q\,)/M.
\end{equation}
Performing the Fourier transformation to coordinate space, one finds the gravitational potential from the massive gauge boson in Yukawa form
\begin{equation}
\Phi_{\text{YU}}\left( r\right) = \alpha_B G_N M\,\frac{e^{-\mu r}}{r} . \label{solution}
\end{equation}%
This scenario is similar to the one recently discussed in \cite{J0,J1,J2,J3,J4,J5}. The mass of the vector boson is related to the length scale via $\mu=1/\lambda$. In galactic scales we expect $\lambda$ to be of kpc order.
From these results, we can derive some phenomenological aspects of this potential for the dynamics of galaxies and use it to obtain a cosmological model. For example, we can explain the emergence of dark matter as a consequence of the modification of the law of gravity. To see this, let us consider the condition for the energy density due to the presence of massive vector particles
\begin{eqnarray}
 \rho_{\text{YU}}= \frac{1}{4 \pi
G_N } \nabla^2 \Phi_{\text{YU}} \geq 0,
\end{eqnarray}
where $\rho_{\text{YU}}$ is the effective
energy density due to the presence of vector particles.
According to the coupling in the interaction Lagrangian in Eq. (\ref{gauge-baryon-coupling}), the $\alpha_B$ parameter should be positive, and we rewrite the potential in Eq. (\ref{solution}) as
\begin{eqnarray}
    \Phi_{\text{YU}}(r)=\frac{\alpha_B G_N M }{r}e^{-\frac{r}{\lambda}}\,,
\end{eqnarray}
which shows that the gravitational force is repulsive. In addition, we have for the energy density
\begin{eqnarray}
 \rho_{\text{YU}}= \frac{M \alpha_B }{4 \pi
r \lambda^2 }e^{-\frac{r}{\lambda}}\geq 0.
\end{eqnarray}

On the other hand, the contribution due to the standard gravitational interaction (massless spin-2 graviton) is given by
\begin{eqnarray}
    \Phi_{\rm B}(r)=-\frac{GM}{r}=-\frac{G_N (1+\alpha)M}{r}\,,
\end{eqnarray}
which instead leads to an attractive force. Therefore, using the rescaling for the
Newton's constant (due to the scalar field) $G = G_N (1+\alpha)$, we find for the total potential
\begin{eqnarray}
   \Phi_{\rm tot}(r)= \Phi_{\rm B}+\Phi_{\text{YU}}=-\frac{G_N M }{r}\left(1+\alpha-\alpha_B e^{-\frac{r}{\lambda}}\right).
\end{eqnarray}
If a point particle with mass $m$ is placed at
distance $r$ in this gravitational potential, it will experience the
following force $\vec{F}=-m \vec{\nabla} \Phi_{\rm tot}(r)$,
yielding
\begin{equation}\label{F5}
F=\frac{ G_N M m }{r^2}
\left[1+\alpha-\alpha_B\left(\frac{r+\lambda}{\lambda}\right)e^{-\frac{r}
{\lambda}}\right]
\end{equation}

Using the fact that
$|\vec{F}|=m\, v^2/r$, we can rewrite the circular speed of an orbiting
test object as
\begin{equation}
v^2 = \frac{ G_N M }{r} \left[1+\alpha-\alpha_B\,\left(\frac{r+\lambda}{\lambda}\right)e^{-\frac{r}{\lambda}}\right].
\end{equation}
Our result for the circular velocity aligns with the those in scalar-vector-tensor gravity \cite{Moffat1,Moffat2,Moffat3,Moffat4}. An interesting result is found for galactic scales, by writing the velocity to get a
MOND-like relation
\begin{equation}
\frac{v^2}{r} = \frac{G_N (1+\alpha) M }{r^2}-\sqrt{\left(\frac{G_N M}{r^2}\right)\left(\frac{G_N M (r+\lambda)^2\alpha_B^2}{r^2 \lambda^2}\right)e^{-\frac{2r}{\lambda}}}.
\label{speed}
\end{equation}
We can define
\begin{eqnarray}
    a_B= \frac{G_N M}{r^2}
\end{eqnarray}
along with
\begin{eqnarray}
    a_{0}= \lim _{r \to \lambda} \frac{G_N M \alpha_B^2 (r+\lambda)^2}{r^2 \lambda^2}e^{-\frac{2r}{\lambda}}=\rm constant,
\end{eqnarray}
which is noted to be constant in the limit $r \to \lambda$.
Although we obtain a characteristic scale for acceleration in our
model, we will now argue that $a_{0}$ in our case is not
precisely the same as the acceleration $a_0$ that appears in MOND theory. In the MOND case, indeed, $a_0$ is a constant, independent of
$a_B$, and cannot be a function of the distance $r$. To see this, we
can define $\mu(a_B/a_{0})={F}/(m a_B)$ and then by expanding
$\mu(x)$, for the MOND theory we should have
\begin{eqnarray}
&&\mu(x)=1  \  \ \  \rm{for}  \ \  \emph{x}\gg 1, \nonumber  \\[2mm]
&&\mu(x)=x \ \  \    \rm{for}  \ \  \emph{x}\ll  1.
\end{eqnarray}
For the total force we get
\begin{eqnarray}
F= m a_B(1+\alpha)-(ma_B) \alpha_B
\left(\frac{r+\lambda}{\lambda}\right)e^{-\frac{r} {\lambda}}.
\end{eqnarray}
Using $F= ma_B \mu(a_B/a_{0})$ in terms of $\mu(a_B/a_{0})$, we
get
\begin{eqnarray}
\mu(a_B/a_{0})=1+\alpha-\alpha_B \left(1+\frac{r}{\lambda}\right)e^{-\frac{r}{\lambda}}.
\end{eqnarray}
Near the galactic center, we have $r \ll \lambda$. If we define $z=r/\lambda$ and expend in series around $z$, we get
\begin{eqnarray}
\mu(a_B/a_{0})=1+\alpha-\alpha_B+\frac{\alpha_B r^2}{2 \lambda^2}+\dots
\end{eqnarray}
The second term encodes corrections arising from Newton's constant, while the other terms come from the modified law of gravity. Basically, these terms play the role of dark matter, which is only an apparent effect in our model and can be neglected in the inner region. An interesting fact is that the extra repulsive force due to $\alpha_B$ (spin-1 graviton) and the extra attractive  force due to $\alpha$ cancel out,  resulting in what MOND predicts,
\begin{eqnarray}
\mu(a_B/a_{0})=1+\alpha-\alpha_B \sim 1,
\end{eqnarray}
provided $\alpha \sim \alpha_B$, and we neglect the term $\alpha_B r^2/\lambda^2$. 
That is a purely Newtonian behavior. 

In the intermediate region when $r \sim \lambda$, we get 
\begin{eqnarray}
\mu(a_B/a_{0})=1+\alpha-\frac{2\alpha_B}{e}
\end{eqnarray}

Finally, in the outer part of the galaxy,  where $r >\lambda$, the repulsive gravitational force
due to spin-1 particles vanishes out since $e^{-\frac{r}{\lambda}}
\to 0$, hence we get a similar constant term as in MOND, but given
by
\begin{eqnarray}
\mu(a_B/a_{0}) \sim 1+\alpha.
\end{eqnarray}
In other words, from the last equations, we see that gravity
appears to be stronger in the outer regions of galaxies not
because of a dark matter particle, but because the repulsive force vanishes, allowing the attractive force to dominate due to the modification of Newton's constant through $\alpha$. The observed effects attributed to dark matter can thus be mimicked by this modified acceleration proportional to $\alpha$, which should be a characteristic of each galaxy. This conclusion is in agreement with the scalar-vector-tensor gravity \cite{Moffat1,Moffat2,Moffat3,Moffat4}. 
\section{Emergence of dark matter in cosmological scales}
\label{EmerCos}
In our setup, we have two type of matter fields: the baryonic matter and the dark energy. In cosmological scales we expect $\lambda$ to be of Gpc order. The cosmological constant in our setup emerges from the condensation of the scalar field, which is, in turn, related to the gauge field. In fact, it determines the mass term of the gauge field. Thus, in general we expect 
\begin{eqnarray}
\mu \sim \Lambda^\delta,
\end{eqnarray}
where, if we set $\delta=1/2$, we can obtain the range of the force
\begin{equation}
\lambda\equiv \frac{1}{\mu}=\dfrac{1}{\sqrt{\Lambda }}\sim 10^{26}\text{ m}\sim \text{Gpc}
\label{lambda_G (Newtonian)},
\end{equation}%
which is the size of the observable Universe and gives the characteristic length scale associated with the dark energy scalar potential. Here $\lambda$ is of  Gpc order and dominates in cosmological scales. As we shall point out, in general, we expect a varying mass $\mu$, which opens the possibility of evolving dark energy.

It should be noted that $\lambda$ is, in principle, expected to differ across different galaxies and length scales. Indeed, we have observed that $\lambda$ is of the order of kpc at galactic scales and Gpc at cosmological scales. One way to explain this is by assuming a mass fluctuations similar to the chameleon mechanism \cite{Khoury:2003rn}. This implies that incorporating matter (with higher density, such as in galaxies) into the dark energy model introduces an interaction that causes $\lambda$ to decrease and $\mu$ to increase, meaning that the mass parameter $\mu$ can vary depending on the environment. On cosmological scales, where the density is lower, the mass is smaller, and thus $\lambda$ increases. Therefore, it follows that
$\lambda (x)=\hslash /(\mu(x) c)$. Then, by differentiating  this equation, we get $\Delta \lambda / \lambda \simeq \Delta \mu / \mu$. 

In order to study the cosmological implications of our model, let us consider a flat, spatially homogeneous and isotropic background spacetime, which is given by the
Friedman-Lemaitre-Robertson-Walker (FLRW) metric
\begin{equation}
ds^2=-dt^2+a^2\left[dr^2+r^2(d\theta^2+\sin^2\theta
d\phi^2)\right], \label{FLRW}
\end{equation}
where we can further use $R=a(t)r$, where $a(t)$ is the time-dependent scale factor, $x^0=t, x^1=r$, and $h_{\mu \nu}$ is the two dimensional metric.  The dynamical apparent
horizon, a marginally trapped surface with vanishing expansion, is
determined by the relation
\begin{equation}
h^{\mu
\nu}(\partial_{\mu}R)\,(\partial_{\nu}R)=0.
\end{equation}

It is important to note that a massive vector gauge boson spin-1 cannot be isotropic. 
At first glance, this may seem to contradict the assumption of a FLRW Universe, which is isotropic. However, this inconsistency can be resolved by introducing a cosmic triad \cite{Armendariz-Picon:2004say,Golovnev:2008cf,Maleknejad:2012fw}. Within this framework, the gauge fields form a mutually orthogonal set of vectors. As a result, the energy-momentum tensor in the spatial directions becomes proportional to the identity matrix, ensuring compatibility with the FLRW metric. For a more detailed analysis and further explanations, we refer to Refs. \cite{Pongkitivanichkul:2020txi,Waeming:2021ytf,Armendariz-Picon:2004say,Golovnev:2008cf,Maleknejad:2012fw}.
We calculate the apparent horizon radius for the FLRW Universe as $R=ar=1/H$,
with $H=\dot{a}/a$ being the Hubble parameter (the overdot denotes derivative with respect to the time $t$. Furthermore, we shall consider $a(0)=1$). 

For the matter source in the FLRW Universe, we shall assume a perfect
fluid described by the stress-energy tensor
\begin{equation}\label{T}
T_{\mu\nu}=(\rho+p)u_{\mu}u_{\nu}+pg_{\mu\nu}, 
\end{equation}
where $\rho$ and $p$ are the energy density and pressure of the fluid, respectively. 
On the other hand, the total mass $M = \rho V$ in the region
enclosed by the boundary $\mathcal S$ is no longer conserved. One can compute the change in the total mass using the thermodynamics law. Furthermore, the conservation equation $\nabla^\mu T_{\mu\nu}=0$
gives the continuity equation $\dot{\rho}+3H(\rho+p)=0$.

Let us now derive the dynamical equation for Newtonian cosmology. Toward this goal, let us consider a compact spatial region $V$ with a compact boundary
$\mathcal S$, which is a sphere having radius $R= a(t)r$, where $r$ does not depend on $t$. Going back and combining the second law of Newton for the test
particle $m$ near the surface with the gravitational force \eqref{F5}, we obtain 
\begin{equation}\label{F6}
\ddot R=\ddot{a}r=-\frac{G_NM}{R^2}\left[1+\alpha -\alpha_B\,\left(\frac{R+\lambda}{\lambda}\right)e^{-\frac{R}{\lambda}}\right].
\end{equation}
This result represents the entropy-corrected dynamical equation for
Newtonian cosmology. 

To derive the modified Friedmann equations of the FLRW Universe in GR, we can use the active gravitational mass $\mathcal M$ rather than the total mass $M$. For the active gravitational mass, we can use the definition
\begin{equation}\label{Int}
\mathcal M =2
\int_V{dV\left(T_{\mu\nu}-\frac{1}{2}Tg_{\mu\nu}\right)u^{\mu}u^{\nu}}.
\end{equation}
If we assume several matter fluids with constant equation of state parameters $\omega_i$ and take the coupling constant to depend on the specific matter source, i.e. $\alpha_i$, we obtain
\begin{equation}\label{addot}
\frac{\ddot{a}}{a} =-\frac{4\pi G_N
}{3}\sum_i \left(\rho_i+3p_i\right)\left[1+\alpha-\alpha_i\left(\frac{R+\lambda}{\lambda}\right)e^{-\frac{R}{\lambda}}\right],
\end{equation}
where the sum runs over all the components that fill the Universe. 
Thus, in general, the corrections due to the Newton constant encoded in $\alpha$ are not the same as the coupling parameters $\alpha_i$. In the following, we will consider two distinct cosmological models.

\subsection{Model I}
Let us first examine the cosmological implications by considering the case when $\lambda \sim R$. In such a case, we get from Eq. \eqref{addot}
\begin{equation}
\frac{\ddot{a}}{a} =-\frac{4\pi G_N
}{3}\sum_i \left(\rho_i+3p_i\right)\left(1+\delta_i\right),
\end{equation}
where we have defined $\delta_i=\alpha-2 \alpha_i/e$. We can obtain the modified equation for the dynamical evolution of the  FLRW Universe written as
\begin{align}
\frac{\ddot{a}}{a} &= - \left(\frac{4 \pi G_N }{3}\right)\sum_i \left(1+3\omega_i\right) \rho_{i 0} a^{-3 (1+\omega_i)} \left(1+\delta_i\right)
\label{2Aaddot1}
\end{align}
where we have used the continuity equation
along with the expression for densities $\rho_i=\rho_{i 0} a^{-3 (1+\omega_i)}$. 

By multiplying $2\dot{a}a$ on both sides and integrating over time, we get
 \begin{align}
    \label{MFE}
  & \frac{\dot{a}^2}{a^2}=  \frac{8\pi  G_N }{3} \sum_i  (1+\delta_i) \rho_{i0} a^{-3 (1+\omega_{i})}\,,
 \end{align}
for a spatially flat Universe, or equivalently
\begin{equation}
H^2 = \frac{8\pi G_N
}{3}\sum_i \rho_i (1+\delta_i)\,. \label{0Fried01}
\end{equation}
Clearly, the standard cosmological model is recovered for $\alpha=\alpha_i=0$.

Now, by making use of critical density $\rho_{\rm crit}=\frac{3}{8 \pi G_N}H_0^2$, we get
\begin{align}\label{eq430}
  E^2(z)\equiv \frac{H^2(z)}{H_0^2}&=\sum_i(1+\delta_i)\Omega_i \,,
\end{align}
where  
\begin{equation}
\label{Ome}
\Omega_i=\Omega_{i0}(1+z)^{3(1+\omega_i)}, \,\,\,\,\, \,\, \Omega_{i0}=  8 \pi G \rho_{i0}/(3H_0^2)\,. 
\end{equation}
In this particular cosmological model, we have three components, namely
\begin{align}\notag
    \sum_i (1+\delta_i) \Omega_i &=(1+\delta_B)\Omega_{B,0} (1+z)^{3}\\
    &+(1+\delta_R)\Omega_{R,0} (1+z)^{4}+(1+\delta_\Lambda)\Omega_{\Lambda,0}\,,
\end{align}
where $z\equiv a^{-1}-1$ is the cosmological redshift, and $\Omega_{B,0}$, $\Omega_{R,0}$ and $\Omega_{\Lambda,0}$ are the density parameters related to baryonic matter, radiation and dark energy, respectively.  The subscript `0' denotes quantities evaluated at present, specifically at $z=0$ where it has been defined.

It is not difficult to see that the dark matter appears as an \emph{apparent effect} from the term
\begin{eqnarray}
    \Omega_{\rm DM} =\delta_B \Omega_{B,0} (1+z)^{3}\,.
\end{eqnarray}
Hence, we can write 
\begin{align}
   E^2(z)&=(\Omega_{B,0}+ \Omega_{\rm DM,0})(1+z)^3+\hat{\Omega}_{R,0} (1+z)^{4}+\hat{\Omega}_{\Lambda,0}
\end{align}
which is effectively the $\Lambda CDM$ model and we have defined $\Omega_{DM,0}=\delta_B\Omega_{B,0}$, $\hat{\Omega}_{R,0}=(1+\delta_R)\Omega_{R,0}$ and $\hat{\Omega}_{\Lambda,0}=(1+\delta_\Lambda)\Omega_{\Lambda,0}$. 
Again, the primary consideration is that dark matter is not a real type of matter in this setup, 
but instead arises due to modifications in the laws of gravity. From the observational measurements, we know that $\Omega_{\rm DM,0}=0.26$ and $\Omega_{B,0}=0.05$, which implies $\delta_B \sim 5.2$.

\subsection{Model II}
Since $\lambda$ is expected to be a very large scale at cosmological distances, it is interesting to study the case where $R\ll\lambda$. In this scenario, by expanding Eq. \eqref{addot} around $x\equiv R/\lambda \ll 1$, we find a modified Friedmann equation of the form
\begin{align}\notag
\frac{\ddot{a}}{a} &= - \left(\frac{4 \pi G_N }{3}\right)\sum_i \left(1+3\omega_i\right) \rho_{i 0} a^{-3 (1+\omega_i)}\\
& \times \left(1+\hat{\alpha}_i+\frac{1}{2}\frac{\alpha_i R^2}{\lambda^2}\right)
\label{2Aaddot}
\end{align}
where $\hat{\alpha}_i\equiv\alpha-\alpha_i$. 
We anticipate that this scenario is the most interesting to explore, as it exhibits the most significant deviations from the standard cosmological model in the high-redshift regime.

Next, by following similar steps to those leading to Eq. \eqref{MFE}, we obtain
 \begin{align}
  & \frac{\dot{a}^2}{a^2}=  \frac{8\pi  G_N }{3} \sum_i  (1+\hat{\alpha}_i) \rho_{i0} a^{-3 (1+\omega_{i})} \nonumber \\
   & -\frac{4 \pi   G_N R^2}{3 \lambda ^2}   \sum_{i} \left(\frac{ 1+ 3 \omega_{i}}{1-3  \omega_{i}} \right) \alpha_i \rho_{i0}  a^{-3 (1+\omega _{i})}.
   \label{58}
 \end{align} 
 From the last equation we get
\begin{equation}
H^2 = \frac{8\pi G_N
}{3}\sum_i \rho_i (1+\hat{\alpha}_i)-\frac{4 \pi G_N R^2}{3} \sum_{i}\Gamma (\omega_i) \rho_i \,,\label{0Fried01bis}
\end{equation}
along with the definitions
\begin{align}
    \Gamma (\omega_i )   & \equiv   \frac{\alpha_i\, (1+3\omega_i)}{  \lambda^2  (1-3\omega_i)}. \label{def-Gamma_2}
\end{align}

It is interesting to see the emergence of an apparent singularity 
in Eqs. \eqref{58}-\eqref{def-Gamma_2}
for $\omega_R=1/3$ (radiation). This adds an intriguing result to our understanding, suggesting a potential phase transition in the early Universe from a radiation-dominated state to a matter-dominated one. One way to resolve this issue is that we expect the interacting parameter between the gravitons and radiation to be small, hence we can set
\begin{eqnarray}
\label{aR}
   \alpha_R = 0.
\end{eqnarray}
This condition resolves the problem of apparent singularity. In simple terms, since $\alpha_R=0$, there is no contribution at all from the radiation. Hence only the coupling term $\alpha_B$ and $\alpha_{\Lambda}$ are important in our setup.
In this way, we can write Eq. \eqref{0Fried01bis} as
\begin{equation}
H^2 = \frac{8\pi G_N
}{3}\sum_i \rho_i (1+\hat{\alpha}_i)-\frac{4 \pi G_N R^2}{3} \sum_{i,\omega_i\neq1/3}\Gamma (\omega_i) \rho_i \,,
\end{equation}
bearing in mind that the definition \eqref{def-Gamma_2} of $\Gamma (\omega_i)$ holds for matter and dark energy, while $\Gamma (\omega_{1/3})=0$ for radiation due to the condition \eqref{aR}.

Next, if we make use of $\rho_{\rm crit}=\frac{3}{8 \pi G_N}H_0^2$, we get the two solutions
\begin{align}
\nonumber
   E^2(z)&=\frac{1}{2}\,\sum_i(1+\hat{\alpha}_i)\Omega_i \\
   &\pm  \frac{\sqrt{(\sum_i (1+\hat{\alpha}_i)\Omega_i )^2-2 \sum_{i} \frac{\Gamma(\omega_i) \Omega_i}{H_0^2}}}{2},
   \label{eq430bis}
\end{align}
with $\Omega_i$ and $\Omega_{i0}$ defined as in Eq. \eqref{Ome}. clearly, we consider only the solution with positive sign as a physical solution, since it recovers the correct behavior in the limit $\alpha=\alpha_i=0$. 

As in the previous model,  we therefore have the following three sources
\begin{align}\notag
    \sum_i (1+\hat{\alpha}_i) \Omega_i &=(1+\hat{\alpha}_B)\Omega_{B,0} (1+z)^{3}\\
    &+(1+\hat{\alpha}_R)\Omega_{R,0} (1+z)^{4}+(1+\hat{\alpha}_\Lambda)\Omega_{\Lambda,0}\,.
\end{align}
Let us argue about the emergence of dark matter in the present setup. Dark matter emerges from the combination of two main terms: the first one arises from the identification
\begin{eqnarray}
    \Omega^{(1)}_{\rm DM}(z)=\hat{\alpha}_B \Omega_{B,0}(1+z)^3.
\end{eqnarray}

On the other hand, the second correction is similar to the one argued in Ref. \cite{J0,J1}, namely if we take the state parameter for matter $\omega_i=0$, then from the term $2  \Gamma|_{\omega_i=0} \Omega_i /H_0^2 $ where $\Omega_i=\Omega_{B,0}(1+z)^{3}$, we can define the following quantity \cite{J0,J1,J2,J3,J4}
  \begin{equation}
   \Omega^{(2)}_{\rm DM}= \frac{1}{\lambda H_0} \sqrt{2 \alpha_B \Omega_{B,0}} (1+z)^{3}.
\end{equation}
The main conceptual difference compared to \cite{J0,J1,J2,J3,J4} is that here dark matter does not appear from a pure Yukawa-like potential that leads to attractive force. In fact, as we saw, the force can be repulsive, which is suppressed by an additional attractive component from Newton’s constant corrections, which in turn can lead to attraction-dominated effect.  Specifically, it has been shown that the dark matter density parameter can be related to the baryonic matter in terms of the relation
\begin{equation}\label{DMCC}
   \Omega_{\rm DM}^{(2)}(z)= \sqrt{2\,\alpha_B\,\Omega_{B,0}  \Omega_{\Lambda,0}}{(1+z)^3}\,,
\end{equation}
where we have defined
\begin{equation}
    \Omega_{\Lambda,0}= \frac{1}{\lambda^2 H^2_0 }. \label{de}
\end{equation}

We aim to incorporate the matter contribution terms that are significant in the early Universe.
Specifically, we include contributions from cold matter ($\omega=0$), radiation ($\omega=1/3$, with $\alpha_R=0$, ensuring no contribution from $\Gamma(1/3)$ and thus avoiding divergence) and dark energy ($\omega=-1$). 
We can write Eq. \eqref{eq430bis} in the form
\begin{equation}
E^2(z)=\frac{1}{2}\left(\mathcal{K}+\sqrt{\mathcal{K}^2-\frac{[\Omega^{(2)}_{\rm DM}]^2}{(1+z)^3}+\alpha_{\Lambda} \Omega_{\Lambda,0}^2}\right),
\label{Esquar}
\end{equation}
where we have defined
\begin{eqnarray}
    \mathcal{K}=(\Omega_{B,0}+\Omega^{(1)}_{\rm DM,0})(1+z)^3+ \bar{\Omega}_{R,0} (1+z)^4+\bar{\Omega}_{\Lambda,0},
\end{eqnarray}
and
\begin{eqnarray}
    \bar{\Omega}_{R,0} &=& (1+\hat{\alpha}_R) \Omega_{R,0},\\[2mm]
    \bar{\Omega}_{\Lambda,0} &=& (1+\hat{\alpha}_\Lambda) \Omega_{\Lambda,0},
\end{eqnarray}
where $\hat{\alpha}_R=\alpha$, due to the assumption $\alpha_R=0$.

As a first case, we consider a specific choice of $\alpha_B$ such that 
\begin{eqnarray}
\Omega^{(1)}_{\rm DM,0}=\Omega^{(2)}_{\rm DM,0},
\end{eqnarray}
which implies
\begin{equation}
\label{abeta}
     \alpha=\alpha_B + \mathcal{F}_{B,\Lambda}\,,
     \end{equation}
where
\begin{equation}
  \mathcal{F}_{B,\Lambda} \equiv \frac{\sqrt{2 \alpha_B \Omega_{\Lambda,0} \Omega_{B,0}}}{\Omega_{B,0}}.
\end{equation}
This means that given $\alpha_B$,  we can get an equation for $\alpha$ by 
\begin{equation}
    \hat\alpha_B=\alpha-\alpha_B = \mathcal{F}_{B,\Lambda},
\end{equation}
and also
\begin{equation}
\label{alambda}
    \hat{\alpha}_{\Lambda}=\alpha-\alpha_{\Lambda}=(\alpha_B-\alpha_{\Lambda})+\mathcal{F}_{B,\Lambda}.
\end{equation}
By substituting Eqs. \eqref{abeta}-\eqref{alambda} into \eqref{Esquar}, we get an expression for $E^2(z)$ that is parametrically dependent on $\alpha_B$ and $\alpha_\Lambda$ only, i.e.
\begin{widetext}
\begin{eqnarray}
\label{NewEq}
    \hspace{-0.7cm}E^2(z)&=&\frac{1}{2}\Bigg\{
\Omega_{B,0}\left(1+F_{B,\Lambda}\right)\left(1+z\right)^3+\Omega_{R,0}\left(1+\alpha_B+F_{B,\Lambda}\right)\left(1+z\right)^4+\Omega_{\Lambda,0}\left(1+\alpha_B-\alpha_\Lambda+F_{B,\Lambda}\right)+\\[2mm]
\nonumber
&&\hspace{-2.5cm}\Big\{
\left[\Omega_{B,0}\left(1+F_{B,\Lambda}\right)\left(1+z\right)^3+\Omega_{R,0}\left(1+\alpha_B+F_{B,\Lambda}\right)\left(1+z\right)^4+\Omega_{\Lambda,0}\left(1+\alpha_B-\alpha_\Lambda+F_{B,\Lambda}\right)
\right]^2
-2\alpha_B\Omega_{B,0}\Omega_{\Lambda,0}\left(1+z\right)^3+\alpha_\Lambda\Omega_{\Lambda,0}^2
\Big\}^{1/2}
    \Bigg\}.
\end{eqnarray}
\end{widetext}

In order to ensure consistency with $\Lambda$CDM at low redshift, we impose the normalization condition $E^2(z = 0) = 1$, which allows us to fix one of the two free parameters - for instance, $\alpha_\Lambda$ - as follows:
\begin{widetext}
   \begin{equation}
    \label{alam}
\alpha_\Lambda=\frac{4\Omega_{B,0}\left(\Omega_{R,0}+\Omega_{\Lambda,0}\right)\mathcal{F}_{B,\Lambda}-2\Omega_{B,0}^2\left(\alpha_B\Omega_{\Lambda,0}-2\right)+4\Omega_{B,0}\left[\Omega_{R,0}+\Omega_{\Lambda,0}+\Omega_{B,0}F_{B,\Lambda}+\alpha_B\left(\Omega_{R,0}+\Omega_{\Lambda,0}\right)-1\right]}{\Omega_{B,0}\,\Omega_{\Lambda,0}\left(4-\Omega_{\Lambda,0}\right)}.
\end{equation} 
\end{widetext}

If we set $\Omega_{R,0}\sim 10^{-5}$, $\Omega_{B,0}=\Omega_{B,0}^{\Lambda \rm CDM}\sim0.05$, $\Omega_{\Lambda,0}=\Omega_{\Lambda,0}^{\Lambda \rm CDM}\sim 0.7$ \cite{ConPlanck}, from the condition $\Omega^{(1)}_{\rm DM,0}=\Omega^{(2)}_{\rm DM,0}\sim 0.26$
along with the choice $\alpha_B\sim1$, we find $\alpha\sim 6.2$ and $\hat{\alpha}_B \sim 5.2$, respectively. Furthermore, from the normalization condition $E(z=0)=1$, we get
\begin{equation}
\label{E0}
    E^2(z=0)=1 \longrightarrow \alpha_{\Lambda} \sim 7.5.
\end{equation}
which, in turn, implies,  $\hat{\alpha}_{\Lambda}=\alpha-\alpha_{\Lambda}\sim -1.3$. 

\begin{figure}[t]
\begin{center}
\includegraphics[width=7.2cm]{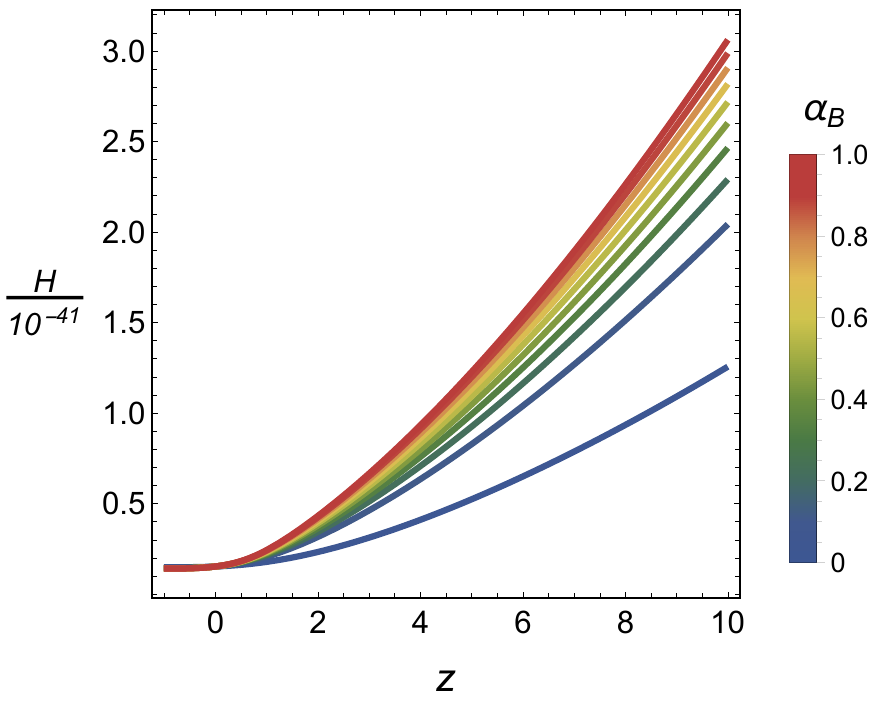}
\caption{Parametric plot of $H(z)$ (in $\mathrm{GeV}$) versus $z$ for
$\alpha_B\in[0,1]$. We set $\Omega_{R,0}\sim 10^{-5}$, $\Omega_{B,0}=\Omega_{B,0}^{\Lambda \rm CDM}\sim0.05$ and $\Omega_{\Lambda,0}=\Omega_{\Lambda,0}^{\Lambda \rm CDM}\sim 0.7$ \cite{ConPlanck}.} 
\label{Pplot}
\end{center}
\end{figure}

The parametric plot of the Hubble rate $H(z)$ versus $z$ is displayed in Fig. \ref{Pplot} for sample values of $\alpha_B$ ranging between 0 and 1. It can be seen that all the curves converge to the current value of $H_0\simeq 10^{-42}\,\mathrm{GeV}$ at present time ($z\rightarrow0$), while they deviate from $\Lambda$CDM ($\alpha_B=0$) at higher redshifts. Specifically, larger values of $\alpha_B$ correspond to higher $H(z)$. As expected, effects of modified gravity in our cosmological setup dominate at higher redshift, leading to a larger value of the Hubble rate compared to the prediction of the standard cosmological model.

Another scenario is the case when $\alpha>>\alpha_B$ and $\alpha>>\alpha_\Lambda$, which correspond to
$\Omega^{(1)}_{\rm DM,0}\gg \Omega^{(2)}_{\rm DM,0}$. In such a case, we may expand Eq. \eqref{Esquar} around $\alpha_B,\alpha_\Lambda<<1$ to get 
\begin{equation}
E^2(z)=\mathcal{K}-\alpha_B \left(\frac{\Omega_{B,0} \Omega_{\Lambda,0}}{2 \mathcal{K}}\right)(1+z)^3+\alpha_{\Lambda} \left( \frac{\Omega_{\Lambda,0}^2}{4\mathcal{K}} \right)+\dots,
\end{equation}
which reduces to $\Lambda$CDM once the subdominant second and third terms are disregarded.

\section{Primordial Gravitational Waves}
\label{PGW} 
Primordial Gravitational Waves (PGWs) are thought to
carry the imprint of quantum fluctuations and potential phase
transitions that occurred during the inflationary phase of the
early Universe \cite{Maggiore:1999vm}. Detecting such signals
would be extremely significant, as it would enable us to explore
the Universe’s history before Big Bang Nucleosynthesis (BBN).
This includes probing phases such as reheating, the hadron and
quark epochs, and early non-standard phases dominated by matter or
kination. Additionally, it would help assess the resulting
implications for the Standard Model of particle physics. Given
that GR is expected to break down in the
ultraviolet (UV) regime due to quantum corrections, the pre-BBN
epoch serves as an ideal setting to test modified gravity theories
related to the early Universe. 

From a theoretical standpoint, GWs
arise from three primary sources classified according to their
mechanisms of generation: astrophysical, cosmological, and
inflationary sources. Astrophysical GWs are significantly
influenced by the mass of the emitting objects, with massive
compact bodies such as black holes expected to generate strong
signals.  For example, with a minimum mass $M\sim M_\odot$, the
maximum frequency is estimated to be around $10\, \mathrm{kHz}$.
Conversely, various cosmological events in the early Universe,
such as first-order phase transitions occurring at approximately
$10^{-2}\,\mathrm{GeV}$  or just below the Grand Unified Theory
scale, can generate GWs with frequencies around
$10^{-5}\,\mathrm{Hz}$ or somewhat below the GHz range,
respectively. Finally, primordial GWs produced during the
inflationary period encompass a frequency range from
$10^{-18}\,\mathrm{Hz}$ to $1\,\mathrm{GHz}$ \cite{Ito:2022rxn}.

In this section, we calculate the PGW spectrum within the
framework of the modified cosmology described by the Model II in Eq. \eqref{NewEq}, considering $\alpha_B$ as a free parameter. To make a comparison with experimental data, we
focus on the range $[10^{-11},10^3]\,\mathrm{Hz}$, which is
expected to be fully tested by current and upcoming GW
observatories. 
We assume that modified gravity primarily affects cosmic evolution and the PGW spectrum at the background level, i.e., through corrections to the Hubble rate. Nevertheless, it is worth emphasizing that additional corrections could arise from the altered behavior of primordial black holes as well. Indeed, the formation and dynamics of such objects is associated with enhanced curvature perturbations at small scales, which can generate a stochastic background of gravitational waves through second-order effects, thereby leaving characteristic imprints on the PGW spectrum (see also~\cite{Basilakos:2023xof,Papanikolaou:2024kjb}).

The behavior of black holes within the framework of Kaluza-Klein (KK) gravity has been recently studied in~\cite{Ju2025}. In particular, for an exact black hole solution surrounded by a massive spin-1 graviton field, it was shown that both the spacetime geometry and the associated physical properties—such as the dynamics of accretion disk matter, quasinormal modes, temperature profiles and differential luminosity—are significantly modified compared to those of a standard Schwarzschild black hole in Einstein gravity. It would therefore be of interest to investigate how such modifications could manifest in the context of primordial black holes. In light of the preliminary results of~\cite{Ju2025}, it is reasonable to expect that, within the framework of our extended gravitational model, distinctive signatures associated with primordial black holes may emerge in the PGW spectrum. A detailed investigation of this possibility, however, goes beyond the scope of the present work and is left for future research.

\subsection{Standard Cosmology}
We begin by examining the features of the PGW spectrum within the standard Cosmology, which will allow us to
define the relic density of primordial gravitational waves,
$\Omega_{GW}$. For this purpose, we basically follow the approach
of \cite{Bernal:2020ywq,Jizba:2024klq,Luciano:2024mcn}. This
foundation will also assist us in establishing the notation. Our
focus will be on tensor perturbations within the flat FLRW background. In this scenario, we
can set the tensor perturbations $h_{00} =  h_{0i} = 0$.
Furthermore, we shall assume the transverse traceless ($TT$)
gauge, i.e. $\partial^i h_{ij}=0$ and $h^i_i=0$.

In this framework, the dynamics of the tensor perturbation in first-order perturbation theory is ruled by~\cite{Watanabe:2006qe}
\be
\label{hdyn}
\ddot h_{ij}+ 3H\dot h_{ij} - \frac{\nabla^2}{a^2}h_{ij} = 16\pi G\hspace{0.3mm} \Pi_{ij}^{TT}\,,
\ee
where $\Pi_{ij}^{TT}$ is the $TT$ anisotropic part of the stress tensor
\be
\Pi_{ij} = \frac{T_{ij}-p\hspace{0.3mm} g_{ij}}{a^2}\,.
\label{TPE}
\ee
Here,  $T_{ij}, g_{ij}$ and $p$ are the stress-energy tensor, the metric tensor and the background pressure, respectively (we have implicitly assumed that latin indexes run over the three spatial coordinates).

The resolution of Eq. \eqref{hdyn} can be simplified by working in Fourier space, where
\be
h_{ij}(t,\vec{x})  =  \sum_{\lambda}\int\frac{d^3k}{\left(2\pi\right)^3}\hspace{0.2mm}h^\lambda(t,\vec{k})\hspace{0.2mm}\epsilon^\lambda_{ij}(\vec{k})\hspace{0.2mm}e^{i\vec{k}\cdot\vec{x}}\,,
\ee
with $\epsilon^\lambda$ being the gravitational tensor obeying $\sum_{ij}\epsilon^\lambda_{ij}\epsilon^{\lambda'*}_{ij}=2\delta^{\lambda\lambda'}$ and $\lambda=+,\times$ are the two independent polarizations (we use the standard convention of denoting three-vectors with arrows over letter symbols).

Following \cite{Watanabe:2006qe}, we can factorize the tensor perturbation $h^\lambda(t,\vec{k})$ as
\be
h^\lambda(t,\vec{k}) = h_{\mathrm{prim}}^\lambda(\vec{k})X(t,k)\,,
\ee
where the transfer function $X(t,k)$
takes into account the
time evolution of the perturbation, while $h_{\mathrm{prim}}^\lambda$ denotes the amplitude of the primordial tensor perturbations. Here we have used the notation $k = |\vec{k}|$.

Using this parameterizations, the tensor power spectrum can be
expressed in the following form~\cite{Bernal:2020ywq} \be
\mathcal{P}_T(k) =
\frac{k^3}{\pi^2}\sum_\lambda\Big|h^\lambda_{\mathrm{prim}}(\vec
k)\Big|^2 =  \frac{2}{\pi^2}\hspace{0.3mm}G\hspace{0.3mm}
H^2\Big|_{k=aH}\,. \ee On the other hand, Eq.~\eqref{hdyn} takes
the form of a damped harmonic oscillator-like equation
\be
X''  +  2\hspace{0.2mm}\frac{a'}{a}X' +  k^2X  =  0\,,
\ee
where the prime denotes derivative with respect to the conformal time $\tau$, such that $d\tau=dt/a$.

Now, the relic density of PGWs resulting from first-order tensor perturbations in the standard cosmological model can be expressed as \cite{Bernal:2020ywq,Watanabe:2006qe}
\begin{eqnarray}
\nonumber
\Omega_{\mathrm{GW}}
(\tau,k)&=&\frac{[X'(\tau,k)]^2}{12a^2(\tau)H^2(\tau)}\,\mathcal{P}_T(k)\\[2mm]
&\simeq&\left[\frac{a_{\mathrm{hc}}}{a(\tau)}\right]^4\left[\frac{H_{\mathrm{hc}}}{H(\tau)}\right]^2\frac{\mathcal{P}_T(k)}{24}\,.
\label{Ttps}
\end{eqnarray}
In the second step, we have taken the average over periods of oscillations, which gives
\be
X'(\tau,k)\simeq  k\hspace{0.2mm} X(\tau,k) \simeq   \frac{k\hspace{0.3mm} a_{\mathrm{hc}}}{\sqrt{2}a(\tau)} \simeq 
\frac{a^2_{\mathrm{hc}}\hspace{0.3mm}H_{\mathrm{hc}}}{\sqrt{2}a(\tau)}\,,
\ee
where $k=2\pi f=a_{\mathrm{hc}}H_{\mathrm{hc}}$ is the wave number scale at the horizon crossing.

\begin{figure}[t]
\begin{center}
\includegraphics[width=8.7cm]{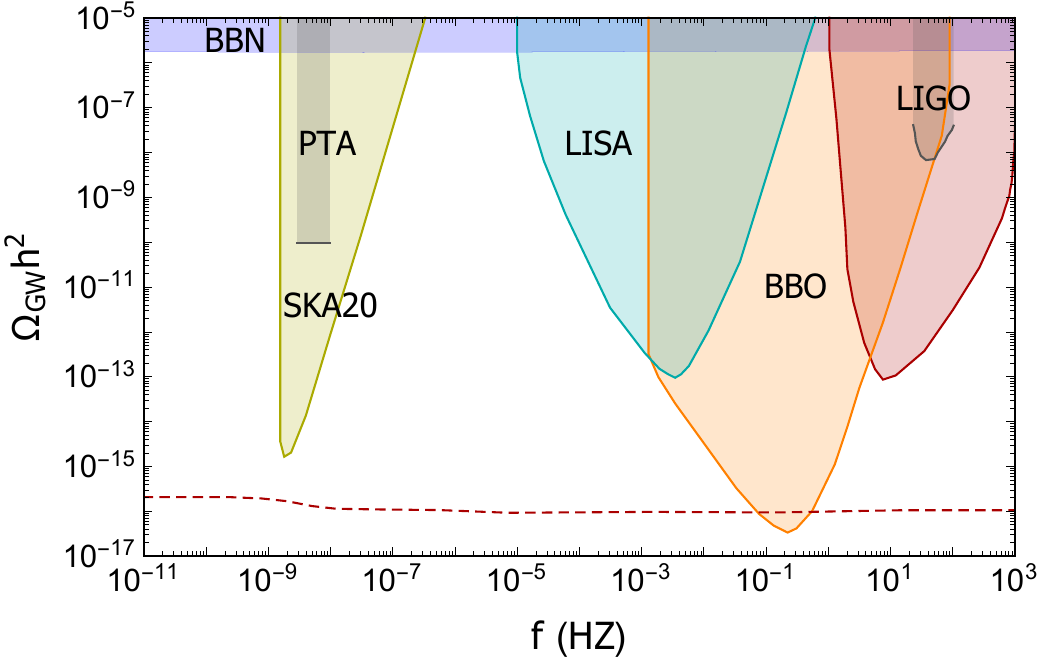}
\caption{Plot of the PGW spectrum versus the frequency $f$ for
$n_T=0$ and $A_S\simeq10^{-9}$, according to standard Cosmology
predictions (dashed red line).} \label{Fig1}
\end{center}
\end{figure}

In turn, the PGW relic density at present time reads
\begin{eqnarray}
&&\Omega_{\mathrm{GW}}(\tau_0,k)h^2 \nonumber \\[2mm]
&&
\simeq\left[\frac{g_*(T_{\mathrm{hc}})}{2}\right]\left[\frac{g_{*s}(T_0)}{g_{*s}(T_{\mathrm{hc}})}\right]^{4/3}\frac{\mathcal{P}_T(k)\Omega_{R}(T_0)h^2}{24}\,,
\label{Spt}
\end{eqnarray}
where $h$ denotes the reduced Hubble constant, while $g_*(T)$ and $g_{*s}(T)$ are the effective numbers of relativistic degrees of freedom that contribute to the radiation energy
density $\rho_r$ and entropy density $s_r$, respectively.
They are defined by $\rho_r=\pi^2 g_*(T)T^4/30$ and $s_r=2\pi^2 g_{*s}(T)T^3/45$.

The scale dependence of the tensor power spectrum is given by
\be
\mathcal{P}_T(k) = A_T\left(\frac{k}{\tilde k}\right)^{n_T}\,,
\ee
where $n_T$ and $\tilde k=0.05\,\mathrm{Mpc}^{-1}$ are the tensor spectral index and a characteristic wave number scale, respectively. Moreover, the amplitude $A_T$
of tensor perturbations is related to the amplitude $A_S$ of scalar perturbations by $A_T=r A_S$, where $r$ is the tensor-to-scalar ratio.

In Fig.~\ref{Fig1}, we have plotted the spectrum~\eqref{Spt}
against the frequency $f$ (dashed red line), setting $n_T=0$ and
$A_S\simeq10^{-9}$ consistently with the Planck observational
constraint at the CMB scale~\cite{ConPlanck}. The colored regions
indicated the projected sensitivities for several GW
observatories~\cite{Breitbach:2018ddu}, including constraints from
LISA interferometer~\cite{LISA:2017pwj}, Einstein Telescope (ET)
detector~\cite{Sathyaprakash:2012jk}, Big Bang Observer
(BBO)~\cite{Crowder:2005nr} and Square Kilometre Array (SKA)
telescope~\cite{Janssen:2014dka}. Furthermore, the BBN bound
arises from the constraint on the effective number of
neutrinos~\cite{Boyle:2007zx,Stewart:2007fu}, while the regions in
gray are those excluded by PTA \cite{KAGRA:2021kbb} and LIGO \cite{Shannon:2015ect}, respectively.

\subsection{Modified Cosmology}
Next, we examine how the modified cosmological scenario in Model II impacts the PGW spectrum. Specifically, we explore the indirect effects on spin-2 gravitational waves by incorporating corrections in the modified Hubble rate described in Eq. \eqref{NewEq}. While the development of a comprehensive formalism for gravitational waves arising from spin-1 gravitons is deferred to future work, our preliminary analysis reveals that the modified cosmological framework may have significant implications for the PGW spectrum. These findings open up the possibility of testing our model through observations by upcoming GW observatories. We also assume that modified gravity mainly influences cosmic evolution at the background
level. In other words, all possible corrections arising from our extended cosmological scenario are incorporated into the modified expression of the Hubble parameter. This holds reasonably well as long as one considers small deviations from GR, which is indeed the specific regime explored in this study. More generally, effects of modified gravity could also manifest at the level of linear perturbations, with a non-trivial impact on the transfer
function and power spectra of both scalar and tensor perturbations generated during the inflationary epoch \cite{Lewis:1999bs, Dival}. However, the exploration of such effects goes beyond the scope of this analysis and will be conducted elsewhere. 

\begin{figure}[t]
\begin{center}
\includegraphics[width=8.7cm]{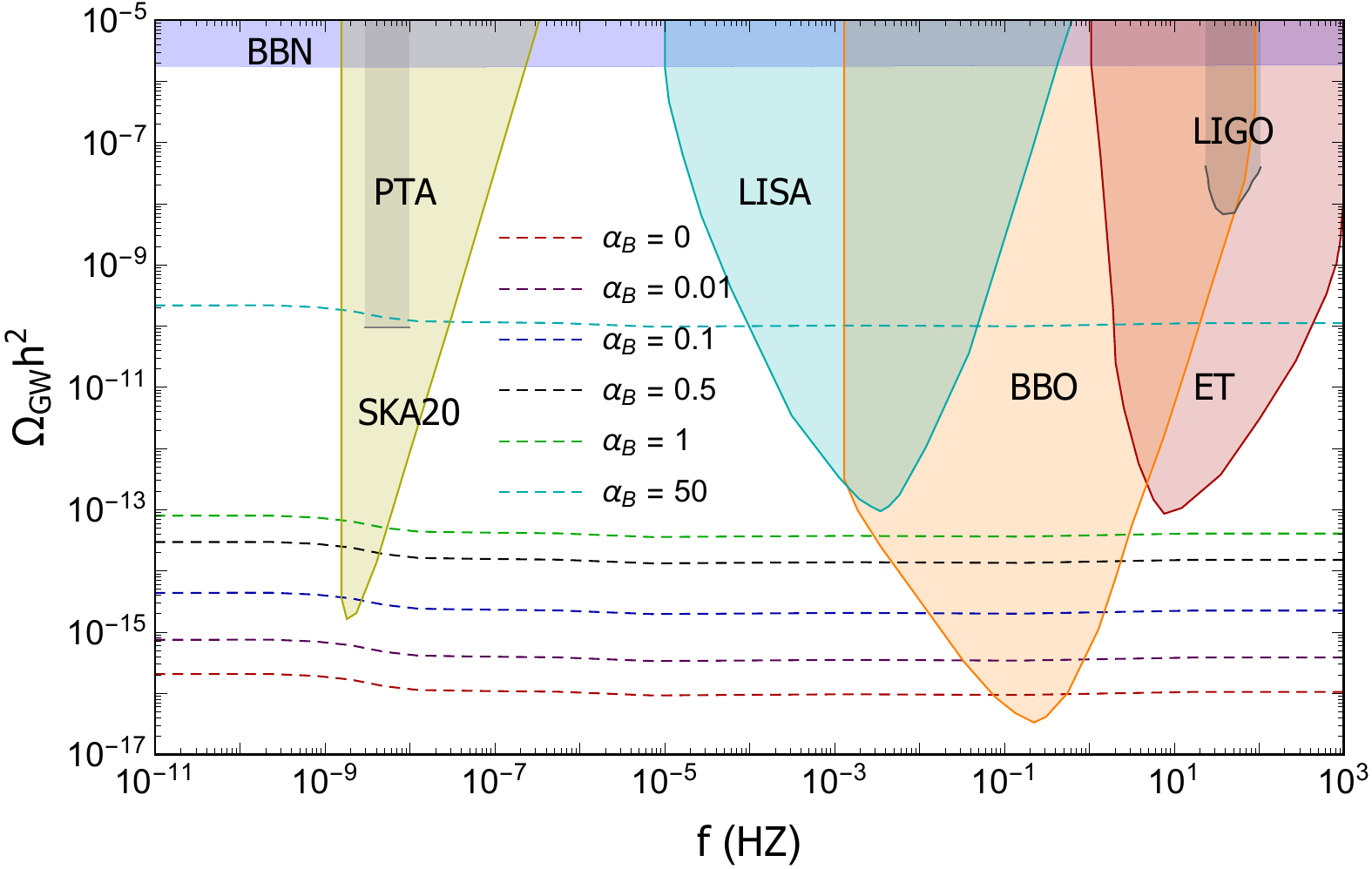}
\caption{Plot of the modified PGW spectrum versus the frequency $f$ for
$n_T=0$, $A_S\simeq10^{-9}$ and several values of $\alpha_B$. } 
\label{Fig2}
\end{center}
\end{figure}

As a first step, we observe that Eq.
\eqref{Ttps} can be equivalently rewritten as
\begin{eqnarray}
&& \Omega_{\mathrm{GW}}(\tau,k)\simeq \left[\frac{a_{\mathrm{hc}}}{a(\tau)}\right]^4\left[\frac{H_{\mathrm{hc}}}{H_{\mathrm{GR}}(\tau)}\right]^2\left[\frac{H_{\mathrm{GR}}(\tau)}{H(\tau)}\right]^2\frac{\mathcal{P}_T(k)}{24} \nonumber \\[2mm]
\nonumber
&&=\,\Omega^{\mathrm{GR}}_{\mathrm{GW}}(\tau,k)\left[\frac{H_{\mathrm{GR}}(\tau)}{H(\tau)}\right]^2\left[ \frac{a_{\mathrm{hc}}}{a_{\mathrm{hc}}^{\mathrm{GR}}}\right]^4 \left[ \frac{a^{\mathrm{GR}}(\tau)}{a(\tau)}\right]^4\left[ \frac{H_{\mathrm{hc}}}{H_{\mathrm{hc}}^{\mathrm{GR}}}\right]^2   \,, \\
\label{eq:PGWBar0}
\end{eqnarray}
where the subscript/superscript ``GR'' indicates the values of the corresponding quantities as defined in traditional GR. It is easy to
check that, for $a(\tau)=a_{GR}(\tau)$, Eq. \eqref{eq:PGWBar0}
gives
$\Omega_{\mathrm{GW}}(\tau,k)=\Omega^{\mathrm{GR}}_{\mathrm{GW}}(\tau,k)$,
which is indeed the standard spectrum Eq.~(\ref{Ttps}) in the
notation we have established.

Using the normalization condition $E(0)=1$, we can write
\begin{equation}
 \Omega_{\mathrm{GW}}(\tau_0,k)
\ \simeq \
\Omega^{\mathrm{GR}}_{\mathrm{GW}}(\tau_0,k)\left[\frac{a_{\mathrm{hc}}}{a_{\mathrm{hc}}^{\mathrm{GR}}}\right]^4\left[
\frac{H_{\mathrm{hc}}}{H_{\mathrm{hc}}^{\mathrm{GR}}}\right]^2
\,. \label{eq:PGWBar}
\end{equation}

\begin{figure}[t]
\begin{center}
\includegraphics[width=7cm]{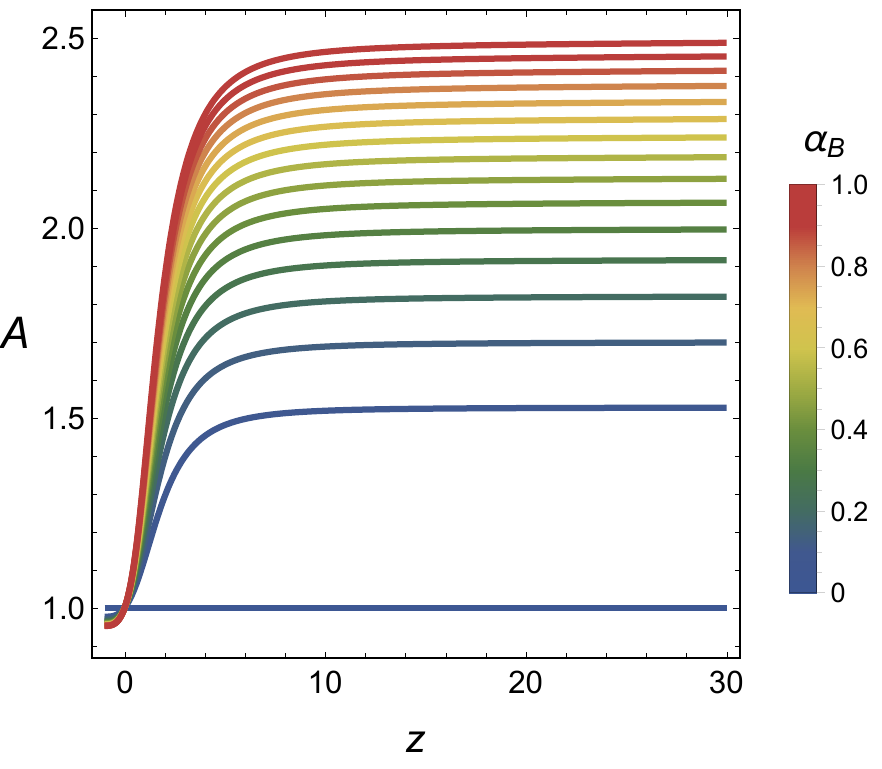}
\caption{Parametric plot of the amplification factor $A(z)$ versus $z$ for
$\alpha_B\in[0,1]$. We set $\Omega_{R,0}\sim 10^{-5}$, $\Omega_{B,0}=\Omega_{B,0}^{\Lambda \rm CDM}\sim0.05$ and $\Omega_{\Lambda,0}=\Omega_{\Lambda,0}^{\Lambda \rm CDM}\sim 0.7$ \cite{ConPlanck}.} 
\label{Aplot}
\end{center}
\end{figure}

The spectrum \eqref{eq:PGWBar} is plotted in
Fig. \ref{Fig2} for several values of $\alpha_B$, displaying an overall shift relative to the standard curve. 
This behavior can be understood by looking at the evolution of the amplification factor $A(z)\equiv \dfrac{H(z)}{H^{GR}(z)}=\dfrac{H(z)}{H(z)|_{\alpha_B=0}}$. 
From Eq. \eqref{NewEq} along with the condition \eqref{alam}, one can check that
$A(z)\rightarrow1$ at present time, while it is nearly constant and higher than unity for the values of $z$ corresponding to the frequency range in Fig. \ref{Fig2} (see Fig. \ref{Aplot}). This implies that 
the PGW spectrum is  not distorted, just showing an overall enhancement of $A^2$.

In passing, we note that a similar modification of the PGW spectrum is also predicted by cosmological models featuring a large number of additional relativistic degrees of freedom in the thermal plasma~\cite{Catena:2009tm}, exotic reheating mechanisms~\cite{Garcia-Bellido:2007fiu,Reh} and early dark energy models~\cite{Zhumabek:2023wka}, as recently emphasized in~\cite{Bernal:2020ywq} (see also \cite{Papan} for the analysis of PGWs from non-canonical inflation).

From Fig. \ref{Fig2}, we see that, for values of $\alpha_B \gtrsim \mathcal{O}(10^{-1})$, signatures of PGWs could potentially be observed by SKA20 (in addition to BBO), even at frequencies below $10^3\, \mathrm{Hz}$. Therefore, should such a global enhancement be detected by future gravitational wave observatories, it could be attributed to a modified expansion rate in the early Universe, as described in Eq.~\eqref{NewEq}.

We would also like to point out that, as expected, relatively large values of $\alpha_B$ (i.e., $\gtrsim \mathcal{O}(10)$) can be excluded, since in this case the relic density would intersect the shadowed region ruled out by PTA measurements (see Fig.~\ref{Fig2}). Moreover, recent constraints from black hole shadow observations in the context of KK gravity~\cite{Ju2025} have provided an upper bound on the parameter $\gamma\equiv \alpha_B / (1 + \alpha)$, namely $\gamma < 0.88$. For the numerical values $\alpha_B \sim \mathcal{O}(1)$ and $\alpha = 6.2$ (see above Eq.~(84)) used in the present analysis, we obtain $\gamma \approx 0.14$, which lies well within the allowed range derived from black hole shadow constraints.

It is worth noting that, more generally, one might expect the values of $\alpha_B$ and $\alpha$ to be context-dependent, potentially varying depending on the specific type of observation. In particular, different observational probes—such as black hole shadows and cosmological data—may yield distinct constraints on these parameters. 

Therefore, in light of the above considerations and results, we can conclude that our extended model exhibits a more intricate and potentially richer phenomenology compared to the standard framework, making it a viable candidate for describing the underlying dynamics of the Universe. Clearly, a comprehensive observational analysis would be both valuable and essential, involving a comparison of the model's predictions with data from Type Ia Supernovae (SNIa), Baryon Acoustic Oscillations (BAO), the Cosmic Microwave Background (CMB) and Cosmic Chronometers (CC), as well as Large Scale Structure observations, such as the growth rate parameter $f\sigma_8$. Such an investigation would help constrain the model's parameters and represents a natural continuation of the present work.

\section{Conclusions}
\label{Conc}
This work investigates the potential of
Kaluza-Klein (KK) gravity to provide insights into the dark sector
of the Universe. By employing KK dimensional reduction, we
reformulate the 5D KK gravity in terms of a 4D spacetime metric
accompanied by additional scalar and vector fields. This framework
suggests the existence of a spectrum of particle states, including
KK gravitons with massive spin-0 and spin-1 states, alongside the
familiar massless spin-2 gravitons of general relativity (GR).  By drawing an analogy with superconductivity, we introduce an addtional complex scalar field and a minimal coupling between this field and the gauge field. We demonstrate that a mass term arises for spin-1 gravitons (gauge field), leading to long-range modifications of
gravity in the form of Yukawa-type corrections. In total, we have seven degrees of freedom. In the gravity sector, there are five degrees of freedom due to the massless graviton in 5D, plus two degrees of freedom in the matter sector, given by the complex scalar field. After dimensional reduction, we obtain two degrees of freedom corresponding to the massless spin-2 graviton, three degrees of freedom for the massive spin-1 graviton, one degree of freedom for the massless spin-0 graviton, and one remaining degree of freedom from the scalar field. In fact, the last degree of freedom from the scalar field can play the role of a dark energy particle and could be a hint of new physics.

This approach effectively links KK theory to scalar-vector-tensor gravity theories (or Moffat's gravity theory), providing a novel perspective on the role of modified gravity in the dynamics of the dark sector.

In addition to the above, we explored the phenomenological aspects of our model. Besides the effects of GR, we have an increase of the gravity force due to the corrections to Newton's constant, along with a repulsive force due to the spin-1 graviton.  This implies that near the galactic center the repulsive force from this spin-1 graviton is suppressed by an additional attractive component from Newton’s constant corrections, resulting in a
Newtonian-like and  attraction-dominated force. In the galaxy’s outer regions, the repulsive force fades
due to its short range, making dark matter appear only as an effective outcome of the dominant attractive corrections. 

Assuming an environment-dependent mass for the spin-1 graviton we obtained the modified Freedman equations and pointed out the emergence of dark matter in cosmological scales. Again, there is no fundamentally particle needed for dark matter in our model, dark matter simply follows from the modified law of gravity as an apparent effect. On the other hand, dark energy plays a fundamental role and  is related to the condensation of the scaler field.

In the end we have studied the effect of dark matter on the PGW spectrum. It is shown that PGW spectrum is sensitive to the apparent dark matter effects which offers the possibility to fully test this theory by future GW observations.

Further aspects are yet to be considered: on a more theoretical note, it is important to investigate whether the presence of a massive spin-1 graviton could give rise to potential issues such as ghosts or instabilities within the theory. As a preliminary remark, we note that the effects of a spin-1 graviton have been shown to lead to stable solutions, at least within the context of black hole physics. Specifically, in the strong gravity regime - such as that associated with black holes - it is necessary to include an interaction term in the energy-momentum tensor. Such configurations have, in fact, been demonstrated to be stable under scalar field perturbations~\cite{Ju2025}. 
Clearly, a more detailed and general analysis requires solving the modified Proca equation within the framework of our extended theory.

On the other hand, from an observational perspective, we intend to investigate the impact of the modified expansion rate, as given by Eq.~\eqref{NewEq}, on the growth of small perturbations and the formation of large-scale structure (LSS) in the early Universe. Indeed, the rate at which LSS evolves from small density fluctuations provides one of the most powerful tests for distinguishing between different cosmological models~\cite{MatPe} (see also~\cite{Mat2,Mat3,Mat4,Mat5} for recent analyses). Specifically, 
one of our next objectives is to examine whether our model can help alleviate the well-known $\sigma_8$ tension. This tension refers to the discrepancy between the amplitude of matter density fluctuations, as inferred from early-Universe observations (such as Cosmic Microwave Background (CMB) measurements by \emph{Planck}), and the lower values obtained from late-Universe probes, including weak lensing and galaxy clustering data. 
Given the non-trivial impact of our modified gravitational framework on the Hubble expansion rate, we expect significant consequences for the dynamics of matter perturbation growth. This expectation is in line with recent results obtained in other extended cosmological contexts~\cite{Sheykhi:2022gzb,Basilakos:2023kvk,Luciano:2025ezl}. In particular, our aim is to fine-tune the free parameter of the model in such a way as to produce an enhanced friction term and an effective Newton’s constant smaller than the standard one, both of which are known to suppress the growth of matter perturbations and could thus offer a consistent resolution to the $\sigma_8$ tension.

Furthermore, it is well-known that the imprint of dark matter, particularly its interactions and composition, can influence the CMB power spectrum \cite{Hu:2001bc}. In fact, the presence of dark matter may affect the acoustic peaks, the ionization history, and the matter-radiation equality epoch. From this standpoint, it is of particular interest to examine how the present model influences the properties of the CMB in order to test it against the latest Planck data and future CMB experiments.  In the present setup, the imprints of dark matter on the PGW spectrum that follows from the propagation of massless spin-2 gravitons, tell us that the background effects via the gravitational influence that mimic dark matter are strong enough to alter the spectrum.  Clearly, this is only an indirect effect, our results demonstrate its sensitivity is sufficient for observation. On the other hand, the direct effect will be to study the propagation and gravity waves and the corresponding spectrum of the spin-1 graviton field, alongside the spin-2 graviton. The study of the spectrum of spin-1 graviton is outside the scope of our work.

Finally, in a broader context, it is essential to compare the present dark matter model with other models proposed in the literature, in order to highlight its virtues and drawbacks both theoretically and experimentally.

Work along these lines is currently underway and will be further explored in future studies.\\

\acknowledgments {The authors are grateful to the anonymous Referee for his valuable comments on the work. The research of GGL is supported by the postdoctoral funding program of the University of Lleida.
GGL  acknowledges  the contribution  of 
the LISA CosWG and of  COST 
Actions   CA21136 ``Addressing observational tensions in cosmology with 
systematics and fundamental physics (CosmoVerse)'', CA21106 ``COSMIC WISPers 
in the Dark Universe: Theory, astrophysics and experiments'', 
and CA23130 ``Bridging high and low energies in search of quantum gravity 
(BridgeQG)''.
The work of A. Sheykhi is based upon
research funded by Iran National Science Foundation (INSF) under
project No. 4022705. DS is supported by the Fundamental Fund of Khon Kaen University. DS has also received funding support from the National Science, Research and Innovation Fund and supported by Thailand NSRF via PMU-B [grant number B39G670016].}


\end{document}